\documentclass[aps,preprint,amsmath,amssymb,amsfonts,nofootinbib]{revtex4-1}
\usepackage{epsfig}
\usepackage{graphicx}
\usepackage{dcolumn}
\usepackage{bm}
\usepackage{amsthm}
\usepackage{amsmath}
\usepackage{color}
\usepackage{hyperref}

\newcommand{\hg}{{\hat g}}

\begin{document}

\title{Turbulence and Random Geometry}
\author{Yaron Oz}
\affiliation{Raymond and Beverly Sackler School of Physics and Astronomy, Tel Aviv University, Tel Aviv 69978, Israel}

\date{\today}
\begin{abstract}

We outline our proposal for a field theory description of steady state incompressible fluid turbulence at the inertial range of scales in a general number
of space dimensions. The theory consists
of a Kolmogorov linear scaling mean field theory dressed by a Nambu-Goldstone dilaton mode that
induces a random measure on the inertial range.
We derive a KPZ-type formula for the anomalous scalings of the velocity structure functions, the velocity gradients
and the local energy dissipation, and relate the dimensionless intermittency parameter to the boundary conformal anomaly.

\vskip 1cm

{\it Dedicated to the memory of Jacob Bekenstein}

\end{abstract}

%\pacs{...}

\maketitle

%\tableofcontents

\section{Introduction}

Most fluid motions
in our universe are turbulent. However, despite centuries of research we still lack
an analytical description and understanding of fluid flows in the non-linear regime.
Turbulence is considered as the most important unsolved problem of classical physics.
The turbulent incompressible fluid flows exhibit  highly complex spatial and temporal intermittent structures. 
When considering their statistical average properties 
a  universal structure is revealed in the inertial range of scales that
exhibits homogeneity, isotropy and anomalous scalings.
Our aim is to gain insight to the nature of the anomalous scalings and construct an analytical formula to calculate them.

In the following we will outline our proposal for a field theory description of steady state incompressible fluid turbulence at the inertial range of scales.
We will consider  a general number of space dimensions $d\geq 2$, with a particular attention being given to the physical two and three space dimensions.
Physically,  our theory consists
of a Kolmogorov linear scaling mean field theory (K41) \cite{Kolmogorov} dressed by a Nambu-Goldstone (NG) gapless dilaton mode. 
The NG mode arises from a spontaneous breaking 
due to the energy flux  of the separate scale and time symmetries of the inviscid Navier-Stokes (NS) equations to a K41 scaling with a dynamical exponent $z=\frac{2}{3}$
\cite{Oz:2017ihc}.
Mathematically, the dilaton mode
induces a random measure on the inertial range and the emerging random geometry structure is reminiscent of two-dimensional
quantum gravity \cite{Knizhnik:1988ak,Distler:1988jt}, with the dilaton being the analog of the Liouville mode. 

We will derive a KPZ-type  \cite{Knizhnik:1988ak,Distler:1988jt} analytical formula for the non-linear anomalous scalings of the velocity structure functions, the velocity gradients
and the local energy dissipation:
\begin{equation}
\Delta - \Delta_{K41} = {\cal G}^2(d) \Delta(1-\Delta) \ , 
\label{KPZg}
\end{equation}
where $\Delta_{K41}$ is the K41 linear scaling dimension of the fluid observable and $\Delta$ is the exact anomalous dimension.
${\cal G}(d)$  is a dimensionless parameter that depends on the number of space dimensions and quantifies intermittency and the deviation from
Kolomogorov linear scaling. It can be measured numerically and experimentally
and we will relate it to the boundary conformal anomaly of the K41 mean field theory \cite{Oz:2017ihc}.
Formula (\ref{KPZg}) can be viewed as a physically and mathematically consistent completion \cite{Eling:2015mxa}
of the random cascade Kolmogorov-Obukhov lognormal model \cite{K62,Obukhov} 
to the strong intermittency regime. It is in agreement with the experimental and numerical data and provides an infinite number
of new predictions  \cite{Eling:2015mxa}.

The inspiration for the proposal comes from classical black hole dynamics.
Black holes have a horizon: objects fall into the black hole through the horizon but cannot emerge. This breaks time reversal
symmetry and allows Einstein equations to describe dissipative effects  (see e.g. \cite{Kovtun:2004de,Bhattacharyya:2008jc}) that are crucial for turbulence.
The evolution equations of the horizon are in the non-relativistic limit the NS equations that describe
incompressible fluid flows \cite{Damour,Eling:2009pb}.
The fluid pressure and velocity denote the deviation from equilibrium of the horizon hypersurface location and
its normal, respectively.
Since the horizon is characterized by a geometry,
the gravity variables suggest that there should be a geometrical framework for studying  the dynamics of fluids:
{\it geometrization of turbulence} (for a brief review see \cite{Eling:2010vr}).

We can map every solution of the incompressible NS equations
to a geometrical configuration of the black hole horizon and its embedding. Hence, studying the statistical properties
of fluid flows  is mapped to studying the statistics of geometries, i.e. {\it random geometry}. 
By a random geometry we will mean  changing the Euclidean measure $d x$ on  $R^d$ to a random measure 
$d\mu (x) \sim  e^{\gamma\phi(x)} d x$,
where the Gaussian random field $\phi(x)$ is log-correlated (for a review see e.g. \cite{chaos}).
We will identify the field $\phi$ with the dilaton and $\gamma$  with the intermittency parameter ${\cal G}$ \cite{Oz:2017ihc}.

The paper is organized as follows. In section II we will construct the field theory of turbulence, while section III will be devoted to a discussion and outlook.

\section{Field Theory of Turbulence}

In this section we construct a field theory of turbulence in the inertial range of scales.
The theory consists of a Kolmogorov linear scaling theory dressed by a NG dilaton
that induces a random measure on the inertial range of scales. 

\subsection{Flux States}

The incompressible NS equations provide a universal description of fluid flows at low Mach number, that is 
$v\ll v_s$ where $v_s$ is the speed of sound.  In flat space they read :
\begin{equation}
\partial_t v^i + v^j\partial_j v^i = -\partial_i p + \nu \partial_{jj} v^i + F^i, ~~~~~~\partial_iv^i = 0,~~~ i=1,...,d  \ ,
\label{NS}
\end{equation}
where $d\geq 2$ is the number of space dimensions, $v^i, i=1...,d$  is the fluid velocity vector, $p$ is the fluid pressure, $\nu$ is the kinematic viscosity and $F^i$ is an external
random force.

In a steady state we have an equality between the incoming energy flux due to the external force and the energy 
dissipation due to viscosity:
\begin{equation}
\int  d^d x \left(F^i v^i\right)  =   \int d^d x  \epsilon(x) \ , 
\label{steady}
\end{equation}
where
\begin{equation}
\epsilon(x) =  \frac{\nu}{2}  \left(\partial_i v^j +  \partial_j v^i \right)^2 \ , \label{epsilon}
\end{equation}
is  the local energy dissipation.

Fluid flows exhibit turbulence at large Reynolds numbers  ${\cal R}_e = \frac{l_c v}{\nu} \simeq 10^3$, where $l_c$
is a characteristic length scale and $v$ is the velocity difference at that scale. In such a state the viscous term $\nu \partial_{jj} v^i$ is negligible compared
to the nonlinear term  $v^j\partial_j v^i$  in the NS equations (\ref{NS}), yet, $\epsilon(x)$ is non-vanishing and balances the incoming flux (\ref{steady}).
This is a turbulent flux state. It is far from equilibrium
and the Gibbs
measure is inappropriate for quantifying its statistics. We will essentially work in the infinite Reynolds number limit.

One defines the inertial range to be the range of distance scales $l \ll r \ll L$,  where the scales $l$ and $L$ are
determined by the viscosity and forcing, respectively.
Although the turbulent flow breaks all the symmetries of the NS equations, 
turbulence at the inertial range of scales  exhibits statistical homogeneity, isotropy and anomalous scalings.
We will be interested in the energy cascade steady state statistics and the equal time correlation functions.

\subsection{Spontaneous Symmetry Breaking of Space and Time Dilatations}

In the absence of a viscosity term, the (inviscid) NS equations (\ref{NS}) exhibit two scale symmetries. These are independent scalings
of space and time which we denote by $R_x\times R_t$:
\begin{equation}
R_x: x^i \rightarrow e^{\sigma}x^i \ , ~~~~~ R_t: t \rightarrow e^{ z \sigma} t \ , \label{2}
\end{equation}
where $\sigma$ and $z$ are  real parameters and
an appropriate charge is being assigned to the random force.
For any value of $z$, one can define the total dimension of an object that carries dimensions $(r_x,r_t)$ under $R_x\times R_t$ as 
\begin{equation}
\Delta_z = r_x + z r_t \ . \label{Deltaz}
\end{equation}
The dimension of $v^i$ is $\Delta_z=1-z$ and the dimension of the  different terms  in the inviscid NS equations is $\Delta_z = 1-2z$.

The local energy dissipation (alternatively the flux) breaks spontaneously the symmetries  of the inviscid NS equations to $R_{z=\frac{2}{3}}$, as follows 
from the requirement that 
\begin{equation}
\Delta_z(\epsilon) = 2-3z = 0 \ .
\end{equation}
This is the Kolmogorov linear scaling $\Delta_{z= \frac{2}{3}}  \equiv  \Delta_{K41}$:
\begin{equation}
 \Delta_{K41}[v^i] = \frac{1}{3} \ .
 \end{equation}
  Thus, we interpret the turbulent flux state
as a state that  spontaneously breaks  the independent scalings
of space and time to $z=\frac{2}{3}$.

When a scale symmetry is spontaneously broken one expects a Nambu-Goldston gapless mode called the dilaton.
We denote the expectation value  (VEV) that breaks $R_x\times R_t \rightarrow R_{z=\frac{2}{3}}$ by:
\begin{equation}
\langle \epsilon \rangle = \bar{\epsilon} \ ,
\end{equation}
where $\langle ...\rangle$ means integration over both space 
and the probability distribution function of the force $F$. 
When considering turbulence in flat space $\bar{\epsilon}$ is a constant. It will be space dependent when studying
turbulent flows in curved spaces.
 
The dilaton $\tau(x)$ 
is the fluctuation:
\begin{equation}
\epsilon(x) = \bar{\epsilon} e^{ \delta \tau(x)}  \label{mode} \ ,
\end{equation}
where $\delta(d)$ is a dimension dependent constant that will be related to intermittency and
$\delta = 0$ means zero intermittency.

The dilaton is invariant under the unbroken $R_{z=\frac{2}{3}}$ symmetry \cite{Hason:2017lao}:
\begin{equation}
\tau(x) \rightarrow \tau(x) + \alpha \sigma + \frac{2}{3}\beta\sigma,~~~~\alpha +  \frac{2}{3}\beta = 0 \ ,
\end{equation}
and $\Delta_{z= \frac{2}{3}} [e^{\tau}] = 0$.
We will denote its charges under $R_x\times R_t$ as $(-Q, \frac{3}{2}Q)$. We will relate $Q$ to the intermittency.

\subsection{The Dilaton Field Theory}

The steady state dilaton effective action in the inertial range of scales should be invariant under the symmetry (\ref{2}) 
and under Galilean boosts which 
forbid time derivative terms in the Lagrangian.
The dilaton action reads \cite{Oz:2017ihc}:
\begin{equation}
S_D(\tau,\hat{g}) = \frac{d}{\Omega_d (d-1)!} \int_M d^d x \sqrt{\hat{g}} \left(\tau {\cal P}_{\hat g} \tau + 2 Q {\cal Q}_{\hat g} \tau \right) \ , \label{Dil}
\end{equation}
where we use the notation of \cite{Levy:2018bdc}.
When considering turbulence in flat space,  $M=B_d(L)$ is the 
$d$-dimensional closed ball in Euclidean space $R^d$ with radius $L$, whose boundary is
the $(d-1)$-sphere $S^{d-1}$ of radius $L$ and $\hat{g}$ is the Euclidean metric.
This will be the case most interesting to us.

$Q$ is a dimensionless parameter called a background charge and $-Q$ is the dilaton charge
under scaling of space as defined above.  
$\Omega_d = \frac{2 \pi^{\frac{d+1}{2}}}{\Gamma[\frac{d+1}{2}]}$ is the surface volume of the $d$-dimensional sphere $S^d$.

${\cal P}_{\hat g}$ are the conformally covariant GJMS operators \cite{GJMS}:
\begin{equation}
\mathcal{P}_{\hat g} =  (-\Box)^{\frac{d}{2}} + lower~ order \ ,
\end{equation} 
where $\Box = {\hat g}^{ab}\nabla_a\nabla_b$ with $\nabla_a$ being the covariant derivative on $M$ and is the partial derivative on $M=B_d(L)$ whose
metric is flat.
They are local in even number of dimensions, for instance
$\mathcal{P}_{d=2}   = -\Box$,
and non-local in odd number dimensions \cite{odd}.
The dilaton is a log-correlated field reminiscent of the Liouville field in two-dimensional quantum gravity.
$\mathcal{P}_{\hat g}$ gets a contribution from the boundary that depends on the boundary conditions.

$\mathcal{Q}_{\hat g}$ is the ${\cal Q}$-curvature  scalar \cite{Q}:
\begin{equation}
 {\cal Q}_{\hat g} = \frac{1}{2(d-1)}(-\Box)^{\frac{d}{2}-1} R + ...  \ .
 \end{equation}
 where $R$ is the scalar curvature with respect to ${\hat g}$.
It is local in even number of dimensions, for instance
${\cal Q}_{d=2} = \frac{1}{2} R$,
and non-local in odd number of dimensions \cite{odd}. $\mathcal{Q}_{\hat g}$ gets a contribution from the boundary, for instance ${\cal Q}_{d=2} = \frac{1}{2} R + K$
where $K$ is the extrinsic curvature. The bulk terms in $\mathcal{Q}_{\hat g}$  vanish on the ball $B_d(L)$ 
and it is determined by the boundary terms.

On conformally flat manifolds, such as the ball, ${\cal Q}_{\hat g}$ is related to the Euler density $E_d$:
\begin{equation}
\int_M d^d x \sqrt{{\hat g}} {\cal Q}_{\hat g} = \frac{1}{2}\Omega_d (d-1)! \chi(M) \ ,
\end{equation}
where $\chi(M)$ is the Euler characteristic of $M$, $\chi(B_d)=1$, and 
we normalized the Euler density as:
\begin{equation}
\int_{S^d} \sqrt{g} d^d x E_d = \Omega_d d! \ .
\end{equation}

Under a Weyl transformation $\hg_{ab} \rightarrow e^{2\sigma}\hg_{ab}$:
\begin{equation}
{\cal P}_{e^{2\sigma}{\hat g}} =  e^{-d \sigma}{\cal P}_{\hat g},~~~~~ {\cal Q}_{e^{2\sigma}{\hat g}}  =  e^{-d \sigma}({\cal Q}_{\hat g} + {\cal P}_{\hat g} \sigma ) \ ,
\end{equation}
and this relation holds both for even and odd number of dimensions.
We have:
\begin{equation}
S_D(\tau - Q\sigma,e^{2\sigma}{\hat g}) = S_D(\tau,{\hat g}) - S_D(Q\sigma,{\hat g}) =  S_D(\tau,{\hat g}) - \frac{d Q^2}{\Omega_d (d-1)!} S_L(\sigma,{\hat g}) \ ,
\label{class}
\end{equation}
where the Liouville action reads:
\begin{equation}
S_L(\phi,\hg) = \int_M d^d x \sqrt{\hg} \left(\phi {\cal P}_{\hg} \phi + 2 {\cal Q}_{\hg} \phi \right) \ , \label{SL}
\end{equation}
and
\begin{equation} 
S_L(\phi - \phi_0,e^{2\phi_0}\hg) = S_L(\phi,\hg) - S_L(\phi_0,\hg) \ .
\end{equation}
The linear shift in the dilaton under Weyl transformation (\ref{class}) means that it can be viewed as a NG boson of a spontaneous breaking of the Weyl
symmetry by the choice of the fixed metric $\hg$.

The dilaton action is changed by a constant under a constant shift:
\begin{equation}
S_D(\tau + \tau_0,\hg) = S_D(\tau,\hg) + d Q \chi(M) \tau_0\ ,
\end{equation}
which imposes a selection rule in correlation functions.

The semiclassical limit of the dilaton field theory (\ref{Dil}) is the large charge limit
$Q \rightarrow \infty$ and the convenient parametrization is in terms of the rescaled dilaton field $\tau_c = \frac{\tau}{Q}$. This limit will correspond to the zero
intermittency limit of the turbulence field theory.

Consider  $M=B_d(L)$.
The two-point function of the dilaton reads:
\begin{equation}
\langle \tau(x) \tau(y) \rangle  = \frac{1}{d} \log \left(\frac{L}{x-y}\right) + ...  \ ,
\end{equation}
where dots refer to a constant and terms that vanish in the limit $L\rightarrow \infty$.
The dimension of the operator $V_{\alpha}(x) = e^{d \alpha \tau(x)}$  reads:
\begin{equation}
\Delta_{\alpha} = \frac{d}{2}\alpha(Q-\alpha) \ . \label{d}
\end{equation}

The dilaton theory (\ref{Dil}) is a generalization of  the two-dimensional Liouville conformal field theory to higher dimensions \cite{Levy:2018bdc}.

\subsection{Dilaton Dressing and Random Geometry}

Define the random measure on the inertial range of scales $d \mu(x)$:
\begin{equation}
d \mu(x)  = \sqrt{g} d^d x,~~~~\sqrt{g} = \sqrt{\hat{g}} e^{d \gamma \tau} \ ,
\end{equation}
where  $\sqrt{\hat{g}} d^d x$ is the fixed measure, which when considering turbulence in flat space  is the Euclidean measure as noted above.
Physically we identify this random measure with the local energy dissipation (\ref{mode}), with 
 $\sqrt{\hat{g}} = \bar{\epsilon}$ and  $\delta = d \gamma$ which vanishes when $\gamma \rightarrow 0$.

Requiring that $\Delta_{\gamma}=d$ (\ref{d}):
\begin{equation}
\frac{d}{2}\gamma(Q-\gamma)  = d \ ,
\end{equation}
we get:
\begin{equation}
Q = \gamma + \frac{2}{\gamma} \ , \label{Qg}
\end{equation}
and
\begin{equation}
\gamma_{\pm} = \frac{Q \pm \sqrt{Q^2-8}}{2} \ .
\end{equation}
There are two branches $\gamma_{\pm}$. We require  $\gamma$ to be  real, hence  $Q^2 \geq 8$. Without loss of generality
we take $Q\geq 0$. We have:
\begin{equation}
 0 \leq \gamma_{-}^2 \leq 2,~~~~~\gamma_{+}^2  \geq 2,~~~~~\gamma_{-}\gamma_{+} = 2  \ .
\end{equation}
The critical value $\gamma = \sqrt{2}$ corresponds in two-dimensional Liouville quantum gravity to the $c=1$ barrier  \cite{Knizhnik:1988ak,Distler:1988jt}.

The operators in the theory are  K41 operators $O_{K41}$ dressed by a dilaton factor $e^{d \alpha \tau}$:
\begin{equation}
O(x) = e^{d \alpha \tau} O_{K41}(x),~~~~~  \alpha = \gamma (1-\Delta)  \ , \label{alpha}
 \end{equation}
where $d \Delta_{K41}$ is the undressed dimension of $O_{K41}$. 
Note, that both $O_{K41}$ and its dilaton factor dressing are made of the fluid velocity and pressure.
Requiring the conformal weight of $O$ to be $d$ we get:
\begin{equation}
\frac{d}{2}\alpha(Q-\alpha) + d \Delta_{K41} = d \ ,
\end{equation}
which using (\ref{Qg}) and (\ref{alpha}) reads:
\begin{equation}
\Delta - \Delta_{K41}= \frac{\gamma^2}{2} \Delta(1-\Delta) \ .  \label{KPZ}
\end{equation}

The partition of the theory in a fixed volume $V= {\cal L}^d$ with respect to the random metric scales as
${\cal L}^{-2d(1+ \frac{1}{\gamma^2})}$.
The expectation value of the one-point function of the integrated dressed operator (\ref{alpha}) in a fixed volume scales as  ${\cal L}^{d(1-\Delta)}$. $d(1-\Delta)$ is the dimension of 
 a set in the random geometry of the inertial range and  $d(1-\Delta_{K41})$ is the dimension with respect to the fixed Euclidean metric
 on it. In general, the expectation value of the $n$-point function in a fixed volume scales  as  ${\cal L}^{d \sum_{i=1}^n(1-\Delta_i)}$.
 
Note, that  in the field theory  framework the natural units for the dimensions (conformal weights) of operators are inverse length, while when we will be applying
(\ref{KPZ}) to turbulence we will associate dimensions of length to the fluid observables. In the fluid setup there is an IR scale $L$,
which can be used in order to relate the two.

\subsection{Anomalous Scaling}

Formula (\ref{KPZ}) reads as  (\ref{KPZg}) with ${\cal G}^2(d) = \frac{{\gamma}^2}{2}$  that quantifies intermittency and the deviation from
Kolomogorov linear scaling.
The limit of zero intermittency  ${\cal G}^2 \rightarrow 0$ is the semiclassical limit 
of the dilaton theory
$Q \rightarrow \infty$ with  ${\cal G}^2 \sim \frac{2}{Q^2}$.
 $\Delta_{K41}$ is the K41 dimension of the operator and $\Delta$ is the exact anomalous dimension.
 We have:
\begin{equation}
0 \leq  {\cal G}^2_- \leq 1,~~~~~ {\cal G}^2_+  \geq 1,~~~~~{\cal G}_{+} {\cal G}_{-} = 1  \ . \label{branch}
\end{equation}
We will assume that the formula can be used when $\Delta_{K41} > 0$.

Consider the  longitudinal structure functions $S_n$:
\begin{equation}
S_n(r) = \langle (\delta_r v )^n \rangle  \sim r^{\xi_n}  \ , \label{structure}
\end{equation}
where $\delta_r v$ is the longitudinal velocity difference between points separated by a fixed distance
$r=|\vec{r}|$:
\begin{equation}
\delta_r v = \left(\vec{v}(\vec{r},t) - \vec{v}(0,t)\right)\cdot \frac{\vec{r}}{r} \ .
\end{equation} 
The K41 scaling dimension of  $(\delta_r v)^n$ is $\Delta_{K41} = \frac{n}{3}$, thus
\begin{equation}
\xi_n -\frac{n}{3}= {\cal G}^2(d) \xi_n(1-\xi_n)  \ . \label{master}
\end{equation}
The numerical and experimental data in two (inverse cascade) and three space dimensions suggest \cite{Eling:2015mxa}:
\begin{equation}
 {\cal G}^2_{d=2} \sim 0,~~~~  {\cal G}^2_{d=3} \sim 0.16  \ ,
\end{equation}
and implies that we should pick the $-$ branch (\ref{branch}). 
The corresponding charges are $Q^2_{d=3} \sim 17$, and if we take  ${\cal G}^2_{d=2}  \sim 0.01$ we get
 $Q^2_{d=2} \sim 205$. Both may be considered much larger than one, thus intermittency in these cases should be explained to a good accuracy
 by the semiclassical dynamics of the dilaton theory.
 
  Formula (\ref{master}) predicts a generic growth of the anomalous scalings at large $n$:
  \begin{equation}
  \xi_n \sim \sqrt{n},~~~~n \gg 1 \ .
  \end{equation} 
  This is a rather different behaviour compared to other existing proposals for the anomalous scalings.
  Unfortunately, at this point 
 an  accurate experimental or numerical data of high enough moments is not available and we cannot check this prediction. 
 
The anomalous dimensions of the transverse structure functions of the velocity vector field as well
as the mixed cases is the same as that  of the longitudinal structure functions since the transverse and the longitudinal velocity 
increments have the same  K41 scaling dimension.

The normalized local space average of the energy dissipation (\ref{epsilon}) over a $d$-dimensional ball of radius $r$, $B_d(r)$,  centered at a point  
$x$ reads:
\begin{equation}
\epsilon_r(x) = \frac{1}{Vol(B_d(r))} \int_{|x'-x| \leq r} d^d x'  \epsilon(x') \ .  \label{mea}
\end{equation}
%The statistics of $\epsilon_r(x)$ depends on $r$ but is independent of $x$ because of homogeneity. 
Consider the quantity $(r\epsilon_r)^{\frac{1}{3}}$. Its K41 scaling dimension is $\frac{1}{3}$. Formula (\ref{KPZg}) suggests that its
statistics is the same as that of $\delta_rv$. Thus, if
\begin{equation}
\langle \epsilon_r^n \rangle  \sim r^{\zeta_n}  \ , \label{eps}
\end{equation}
we get:
\begin{equation}
\zeta_{\frac{n}{3}} = \xi_{n} - \frac{n}{3}   \ , \label{rel}
\end{equation}
with $\xi_n$ given in (\ref{master}).

Next, we define the quantity $D_r^k \delta_r v$ where $D_r^k = (r \partial_r)^k$.  It has the same K41 scaling dimension
as $\delta_rv$, thus the anomalous dimensions of these operators are: 
\begin{equation}
\langle (D_r^k\delta_r v)^n \rangle  \sim r^{\xi_n}  \ .\label{structureder}
\end{equation}

\subsection{Comparison to Random Cascade Models}

Random cascade models for the redistribution of fluid energy via splitting into finer scale eddies allows a rich phenomenological
framework for studying fluid turbulence and intermittency (for a review see e.g. \cite{Frisch,chaos}). 
When the random variable is lognormally distributed one gets the Kolmogorov-Obukhov lognormal model  \cite{K62,Obukhov}.

At leading order in the intermittency parameter ${\cal G}^2$, $\xi_n$ of (\ref{master}) coincides  with the  lognormal model:
\begin{align}
\xi_n - \frac{n}{3}  = {\cal G}^2 \frac{n}{3} \left(1-\frac{n}{3} \right) \label{KO} \ .
\end{align}
They differ at higher orders in ${\cal G}^2$ and  (\ref{master}) 
may be viewed as a completion of the Kolmogorov-Obukhov formula (\ref{KO}) to the strong intermittency regime.

Unlike (\ref{KO}) which implies physically inconsistent supersonic velocities at large $n$ and a violation
of the convexity inequality,  
its completion (\ref{master}) is mathematically and physically consistent.

In the lognormal model, the moments of the local energy dissipation $\epsilon_r $ scale as:
\begin{equation}
\langle \epsilon_r^n \rangle  \sim r^{{\cal G}^2  n(1-n)}  \ , \label{eps}
\end{equation}
and
\begin{equation}
 \langle (\delta_r v)^n \rangle \sim \langle (r\epsilon_r)^{\frac{n}{3}} \rangle \sim  \langle \epsilon_r^{\frac{n}{3}} \rangle r^{\frac{n}{3}} \sim
 r^{{\cal G}^2  \frac{n}{3}(1-\frac{n}{3}) + \frac{n}{3}}   \ , \label{fact}
\end{equation}
as in (\ref{KO}).
In the random cascade models one separates the dynamics of the K41 theory from the that of the random measure and this is the reason
for the factorization in (\ref{fact}). However, this separation of the dynamics is only an approximation that can be made
at weak intermittency. In our language, the dilaton field interacts with the K41 theory and we cannot seperate their dynamics.

More generally, in the random cascade models with a random variable $W$ one has:
\begin{equation}
\langle \epsilon_r^n \rangle  \sim r^{\zeta_n},~~~~\zeta_n = - \log_2(\langle W^n \rangle)  \ , \label{eps2}
\end{equation}
which coincides with (\ref{eps}) when $W$ is lognormally distributed,
and 
\begin{equation}
\xi_{n} = \frac{n}{3}   - \log_2(\langle W^{\frac{n}{3}}\rangle)   \ . \label{rel2}
\end{equation}
In the random geometry framework we get:
\begin{equation}
\xi_{n} = \frac{n}{3}   - \log_2(\langle W^{\xi_n}\rangle)   \ , \label{3}
\end{equation}
which coincides with (\ref{master}) when $W$ is lognormally distributed.

\subsection{Conformal Anomaly and Intermittency}

On the ball $B_d(L)$
the conformal anomaly coefficient $a$ is obtained 
by varying the field theory partition function $Z_{B_d}$:
\begin{equation}
a \sim \frac{\partial}{\partial \log L} Z_{B_d}(L) \ .
\end{equation}
Another way to see the anomaly is by the non-invariance of the partition function under a Weyl transformation that results in 
a term of the form $e^{-a S_L}$, where $S_L$ is the Liouville action (\ref{SL}) and we neglect terms that are suppressed in the inertial
range \cite{Oz:2017ihc}.
We will require that the total conformal anomaly vanishes:
\begin{equation}
a_{total} = a_{dilaton} + a_{K41} = 0 \ . \label{a}
\end{equation}
We may justify this requirement by asking that the inertial range universal structure and in particular the anomalous scalings
should not depend on $L$.
In an even number of dimensions we have \cite{Dowker:2010bu,Oz:2017ihc,Levy:2018bdc}:
\begin{equation}
a_{dilaton} = \frac{2}{\Omega_{d}(d!)^2}\int_{0}^{\frac{d}{2}} dt \prod_{i=0}^{\frac{d}{2}-1}\left(i^2-t^2\right)  +\frac{(-1)^{\frac{d}{2}}}{\Omega_{d}(d-1)!}Q^2  \ .
\label{a2}
\end{equation}
Since $Q^2 \geq 8$ we see that the first term in (\ref{a2}) can be neglected and 
\begin{equation}
a_{dilaton} \simeq \frac{(-1)^{\frac{d}{2}}}{\Omega_{d}(d-1)!}Q^2  \ ,
\label{ad}
\end{equation}
that is its classical value (\ref{class}).
Using (\ref{a}) and the large $Q$ behaviour of the intermittency parameter ${\cal G}^2 \sim \frac{2}{Q^2}$  we get:
\begin{equation}
{\cal G}^2(d)  \simeq \frac{2} {\Omega_{d}(d-1)! |a_{K41}(d)|} \ . 
\label{ad}
\end{equation} 
This relation holds also for odd number of dimensions, where in both cases the contribution to the anomaly comes from the boundary 
\cite{Graham:1999pm,Oz:2017ihc}.

\section{Discussion and Outlook}

We outlined our proposal for a field theory description of the universal structure of incompressible fluid
turbulence at the inertial range of scales, derived an analytical formula for anomalous scalings and related the
intermittency parameter to the conformal anomaly coefficient. 
Note, that if the K41 theory is not conformally invariant \cite{Oz:2018yaz} the same analysis as above holds but $a_{K41}$  (\ref{ad}) gets 
contributions on the ball not only from the Euler density term.
Mathematically, the structure of the field theory is that of a dynamical system (K41 theory) coupled to a random geometry.
Physically, the random geometry structure is induced by a NG dilaton that is a consequence of a spontaneous symmetry breaking
in the turbulent flux state and the dilaton field is responsible for the intermittency. 
There are infinite number of anomalous scaling dimensions predicted by the theory that may  be checked numerically or experimentally.

There are numerous directions to follow.
A calculation of the anomaly coefficient of the K41 mean field theory $a_{K41}$ will enable via the relation (\ref{ad}) to calculate
the intermittency parameter and compare with numerical and experimental data.
For this we need to gain a better understanding of K41 theory, which is a non-local field theory.
A calculation of equal time correlations  of the local energy dissipation such as $\langle (\epsilon_r(x))^p  (\epsilon_R(x))^q \rangle$,
$\langle (\epsilon_r(x_1))^p  (\epsilon_r(x_2))^q \rangle$
and velocity differences $\langle (\delta_r v)^p  (\delta_R  v)^q \rangle$ (see e.g. \cite{biferale}), where $r\neq R$ are in the inertial range, can shed more light on the theory and may be compared with numerical and experimental data.

The random cascade phenomenological framework 
for studying fluid turbulence and intermittency is not directly related to the NS equations that govern the motion of fluid flows.
The spontaneous symmetry breaking mechanism that we introduced provides a direct link between the two via the dilaton that introduces
the random structure.
This topic deserves further study, in particular since the spontaneous symmetry breaking is in the flux state which is far from equilibrium.

It has been stressed by Polyakov \cite{Polyakov:1992er} that the IR boundary conditions at the forcing scale $L$ are of much importance  for turbulence
and
that IR condensates break spontaneously  conformal symmetry. Our dilaton theory is a conformal field theory \cite{Levy:2018bdc} and
it would be valuable  to understand the IR effects on it as well as on the K41 mean field theory.
In our framework the natural units for dimensions are inverse length and an explicit introduction
of the IR scale $L$ can explain the change to units of length.

A study of the non-local dilaton theories in odd number of dimensions may shed more light on odd-dimensional turbulence in general
and on three-dimensional turbulence in particular.
Also, studying turbulence in curved space may offer a new viewpoint on its dynamics. In our framework it amounts to introducing a random dressing of a fixed curved metric on the inertial range.

Higher-dimensional turbulence is mathematically intriguing and may also provide a $\frac{1}{d}$ expansion scheme. Denote by $u$ the velocity one-form, then $du$ is the vorticity two-form and the 
incompressibility condition reads $d^{*}u=0$. In addition to the energy one has a conserved helicity made from the wedge products
${\cal H} = \int_M (du)^m$ when
${\rm dim} M=2m$ is even and ${\cal H}  = \int_M u\wedge (du)^m$ when $dim M =2m+1$ is odd \cite{arnold}.
Studying the turbulent cascade as a function of the number of space dimensions is likely to provide valuable insights.

One may try to generalize  our framework to other systems. An interesting case is Burgers turbulence,
which will correspond to a one-dimensional random geometry. One may expect that the shock wave solutions of Burges equations
that determine the anomalous scaling correspond to singular one-dimensional manifolds.

Finally, the structure of our proposed field theory of steady state inertial range turbulence is reminiscent of the macroscopic structure of quantum gravity 
as a sum over spacetime geometries. It will be interesting to obtain a better understanding of this relationship.

\section*{Acknowledgements}
I would like to thank L. Biferale, V. Mukhanov and A. Polyakov for discussions.
This work  is supported in part by the I-CORE program of Planning and Budgeting Committee (grant number 1937/12), the US-Israel Binational Science Foundation, GIF and the ISF Center of Excellence.

\end{document}